\documentclass[twocolumn,pra,superscriptaddress,longbibliography,nofootinbib]{revtex4-2}

\usepackage[dvips]{graphicx} % for figures
\usepackage{amsfonts,amscd,amsmath,amsthm}
\usepackage{enumerate}
\usepackage{epsfig}
\usepackage{subfigure}
\usepackage{xcolor}
\usepackage[colorlinks = true]{hyperref}
\usepackage{physics}
\usepackage{epstopdf}
\usepackage{framed}
\usepackage{multirow}
\usepackage{color}
\usepackage{longtable}
\usepackage{comment}
\usepackage[ruled,vlined]{algorithm2e}
\usepackage[most]{tcolorbox}

\graphicspath{{./figure/}}

\usepackage{tikz}
\usetikzlibrary{tikzmark, calc, fit, positioning}
\usetikzlibrary{shapes}

\newtheorem{proposition}{Proposition}

\newtcolorbox[auto counter]{mybox}[2][]{
	enhanced,
	breakable,
	colback=blue!5!white,
	colframe=blue!75!black,
	fonttitle=\bfseries,
	title=Box \thetcbcounter: #2,#1
}

\usetikzlibrary{arrows.meta,shadows.blur}

\newcommand{\mc}{\mathcal}

\definecolor{cool_green}{rgb}{0.0, 0.5, 0.0}

\begin{document}

\title{Playing Nonlocal Games with Little to No Shared Randomness}

\author{Mingze Xu}
\email{mingzexu@illinois.edu}
\author{Eric Chitambar}%
 \email{echitamb@illinois.edu}
\affiliation{Department of Electrical and Computer Engineering,\\ Coordinated Science Laboratory\\ University of Illinois Urbana-Champaign, Urbana, IL, USA}%

\date{\today}% It is always \today, today,

\begin{abstract}
   Bell nonlocality reveals correlations that cannot be explained by classical models and plays a central role in quantum information theory. In this work, we investigate classical models of Bell nonlocality under restrictions on shared randomness. For bipartite scenarios with bounded shared randomness, the set of classical correlations becomes nonconvex. We characterize the correlations achievable without shared randomness through the simultaneous evaluation of multiple linear Bell functionals. We then extend our analysis to quantum networks by relaxing the standard assumption of source independence. In this setting, the feasibility constraints derived for the bipartite case can be used to certify source dependence, and we further construct nonlinear inequalities that distinguish classical models with correlated sources from correlations achievable in standard quantum networks. As an application, we show that these nonlinear network inequalities give rise to an entropic Bell inequality for bipartite scenarios with limited shared randomness.
\end{abstract}

\maketitle

\section{Introduction}

One of the most remarkable features of quantum theory is that it allows for correlations that cannot be explained in the framework of classical physics. 
This phenomenon, first shown by Bell \cite{Bell_1964_EPR}, provides an operational way to witness quantum resources without making any assumption about the internal functioning of the devices used, and many experiments have been carried out to demonstrate these quantum effects \cite{Freedman_1972_Experiment-LHV,Fry_1976_Experimental-LHV,Aspect_1982_Experiment-EPR,Weihs_1998_violations-of-Bell}. 
From a foundational perspective, Bell's theorem has reshaped the way we think about nature, and it has stimulated a line of research aiming at characterizing the sets of correlations that are feasible under different physical models \cite{Fine_1982_HV,Barrett_2005_Nonlocal-correlations,Pironio_2005_Lifting-Bell-ineq,Brunner_2014_Bell-nonlocality,Donohue_2015_idenfy-nonconvexity,Dawei_2025_Quantum-Nonlocality-Latency}. 
Bell nonlocality also serves as a practical resource in a series of tasks, such as device-independent quantum information processing \cite{Acin_2007_DI,Pironio_2016_Focus-on-DI}, quantum key distribution \cite{Ekert_1991_QCrpto,Acin_2006_From-Bell-to-QKD}, and demonstrations of computational power \cite{Anders_2009_computational-power}.

In standard Bell scenarios, classical correlations are described by a local-hidden-variable (LHV) model in which shared randomness is viewed as a free resource.  
That is, no restrictions are placed on how much shared randomness is used in conjunction with local processing by the classical agents. 
With free shared randomness, the physically realizable set of correlations has a convex structure, and therefore linear Bell inequalities suffice to separate classical correlations from nonlocal ones. 
If $p(a,b|x,y)$ describes the input/output correlations in a two-party Bell scenario, then a Bell inequality bounding the set of classical correlations has the general form
\begin{equation}
    \label{Eq:Bell-inequality}
    \sum_{a,b,x,y}\beta_{x,y}^{a,b}\,p(a,b\vert x,y)\leq L.
\end{equation}
Beyond this type of linear constraint on the correlations, entropic inequalities have also been considered to bound the classical set \cite{Braunstein_1988_info-theoretic-Bell-ineq,Cerf_1997_entropic-Bell-ineq,Chaves-Fritz_2012_entropic,Chaves_2014_causal}. 
One appealing feature of the entropic approach is that it naturally enables the entropy of the shared random variable to be incorporated into the structural constraints of the classical set.
%In the LHV model, $\Lambda$ accounts for the pre-shared classical correlations between Alice and Bob. For a fixed value of $\Lambda$, Alice and Bob are independent. In other words, $\Lambda$ can be interpreted as all the available shared randomness between the parties.

In practice, spatially separated agents may not have access to a large supply of shared randomness, and in some cases may have none whatsoever.
Motivated by both and resource-theoretic considerations, it is then interesting to study the consequences of limiting or restricting the available shared randomness in different ways. 
Such restrictions fundamentally alter the structure of the classical correlation set, generally making it nonconvex and precluding a tight characterization via linear Bell inequalities like Eq. \eqref{Eq:Bell-inequality}.
The main goal of this paper is to understand how restricted shared randomness affects the structure of classical correlations in both the two-party Bell scenario as well as in multi-source quantum networks.

Several works have already explored alternative methods for studying the classical set in this direction.
For example, entropic Bell inequalities in \cite{Chaves_2014_causal} were obtained that capture constraints on the entropy of the shared random variable. 
In another direction, covariance Bell inequalities have been proposed in \cite{Pozsgay_2017_cov-Bell-ineq} whose nonlinear structure is able to witness shared randomness.
Other matrix-based methods have been developed that relate feasible classical correlations to the size of the shared classical variable required to generate them \cite{Pal-2009a, Donohue_2015_idenfy-nonconvexity, Sikora-2016a, deVicente-2017a}.

Beyond the standard Bell scenarios, recent progress has extended the study to quantum networks. 
A general quantum network possesses several independent sources that each distribute either classical or quantum states to certain subsets of parties, according to the particular network topology \cite{Fritz_2012_beyond-Bell}. 
This framework presents a number of radically novel phenomena, such as the emergence of quantum nonlocality without inputs \cite{Fritz_2012_beyond-Bell,Elie_2019_inflation,Renou_2019_genuine-Q-triangle,Renou_2022_Nonlocality-for-generic,Pozas_2023_Proofs-network-Q-Nonlocality}. 
Meanwhile, new nonlinear inequalities have been derived to certify nonlocal correlations arising from quantum networks \cite{Luo_2018_computationally-efficient-Bell,Rosset_2016_nonlinear-Bell,Tavakoli_2014_nonlocal-star}. 
These results rely on the assumption that all sources in the network are mutually independent, an assumption that we relax in this paper.

The main contributions of this work are as follows.
For the bipartite Bell scenario without shared randomness, we analyze the nonconvex correlation set by identifying tuples of Bell expressions that cannot be simultaneously attained by any classical model. 
This is carried out in Section \ref{Sec:Bell-noSR}, and it leads naturally to new types of nonlocal games that certify the presence of shared randomness between two parties.
We then consider scenarios in which the parties have access to some shared randomness, but impose restrictions on either its deviation from independence or its entropy.
In Section \ref{Sec:Qnetworks}, we cast this analysis in the more general setting of quantum networks and lift the standard assumption of source independence.
We present modified versions of well-known nonlinear network inequalities \cite{Braciard-2012a, Rosset_2016_nonlinear-Bell,Tavakoli_2014_nonlocal-star} that account for the possibility of correlated sources.
Before proceeding to our results, we begin in Section \ref{Sect:LHV-model} by reviewing the structure of classical LHV models in more detail.

%In Sec.\,\ref{Sec:Qnetworks}, we study the relaxation of this assumption and construct nonlinear inequalities tailored for this situation. Specifically, in the bipartite case, these nonlinear network inequalities yield a Bell inequality under limited shared randomness.

%In the standard Bell scenario, parties are assumed to possess an unlimited amount of shared randomness, but from a resource-theoretic perspective, it is meaningful to study the case in which the shared randomness is limited, which imposes nonconvex constraints on the feasible correlation set.

%In Sec.\,\ref{Sec:Bell_test}, instead of devising new nonlinear inequalities, we consider the simultaneous evaluation of multiple linear Bell functions and characterize the feasible region when the classical dimension is bounded.

\section{A general model for Bell nonlocality}

\label{Sect:LHV-model} 

\begin{figure}[b]
\centering
\begin{tikzpicture}[
  font=\large,
  >=Latex,
  line cap=round,
  line join=round,
  % Styles
  source/.style={
    circle, draw=black, line width=0.9pt,
    minimum size=12mm,
    fill=white
  },
  party/.style={
    circle, draw=black, line width=0.9pt,
    minimum size=12mm,
    fill=gray!15
  },
  inarrow/.style={
    -{Latex[length=2.5mm,width=2mm]},
    line width=1.0pt
  },
  outarrow/.style={
    -{Latex[length=2.5mm,width=2mm]},
    line width=1.0pt
  },
  hidden/.style={
    dashed, line width=1.0pt
  },
  dots/.style={font=\Large}
]

% -------------------------
% Top row: sources
% -------------------------
\node[source] (S1) at (0.5,3.3) {$S_1$};
\node[source] (S2) at (2.4,3.3) {$S_2$};

% Middle dots and last source Sm
\node[dots] (sdots) at (4.2,3.3) {$\cdots$};
\node[source] (Sm) at (6,3.3) {$S_m$};

% -------------------------
% Bottom row: parties A_1, A_2, ..., A_n
% -------------------------
\node[party] (A1) at (0,0) {$\mathcal{A}_{1}$};
\node[party] (A2) at (2.6,0) {$\mathcal{A}_{2}$};
\node[dots] (adots) at (4.6,0) {$\cdots$};
\node[party] (An) at (6.6,0) {$\mathcal{A}_{n}$};

% -------------------------
% Inputs x_i (slanted arrows into each party)
% -------------------------
\node (x1) at (-0.9,0.9) {$x_{1}$};
\draw[inarrow] (x1) -- ($(A1.north west)+(0.35,0.1)$);

\node (x2) at (1.7,0.9) {$x_{2}$};
\draw[inarrow] (x2) -- ($(A2.north west)+(0.35,0.1)$);

\node (xn) at (5.7,0.9) {$x_{n}$};
\draw[inarrow] (xn) -- ($(An.north west)+(0.35,0.1)$);

% -------------------------
% Outputs a^i (down arrows)
% -------------------------
\node (a1) at (0,-1.35) {$a_{1}$};
\draw[outarrow] (A1.south) -- (a1);

\node (a2) at (2.6,-1.35) {$a_{2}$};
\draw[outarrow] (A2.south) -- (a2);

\node (an) at (6.6,-1.35) {$a_{n}$};
\draw[outarrow] (An.south) -- (an);

% -------------------------
% Hidden variables / shared sources (dashed arrows)
%   Match your sketch: S1 -> A1 and A2; S2 -> A2; Sm -> An
% -------------------------
\draw[hidden,-{Latex[length=2.5mm,width=2mm]}] (S1.south) -- ($(A1.north)+(0,0.05)$);
\draw[hidden,-{Latex[length=2.5mm,width=2mm]}] (S1.south) -- ($(A2.north)+(-0.1,0.05)$);

\draw[hidden,-{Latex[length=2.5mm,width=2mm]}] (S2.south) -- ($(A2.north)+(0.1,0.05)$);

\draw[hidden,-{Latex[length=2.5mm,width=2mm]}] (Sm.south) -- ($(An.north)+(0,0.05)$);

\end{tikzpicture}
\caption{A general network consists of $m$ sources distributing physical systems to $n$ parties. Each party $i$ receives an input $x_i$ and gives an output $a_i$.}
\label{Fig:general_network}
\end{figure}
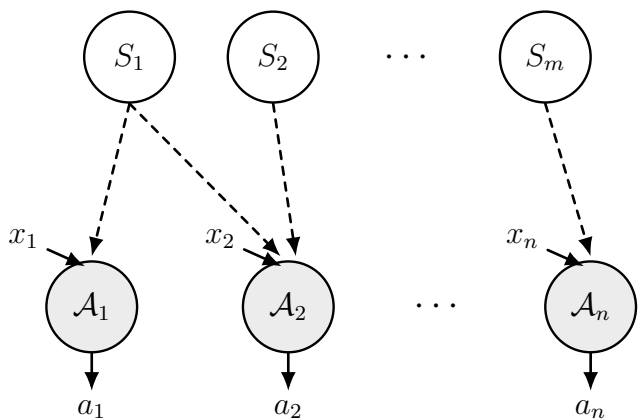

A general model for studying multipartite Bell nonlocality involves $m$ sources, denoted as $S_1,S_2,\cdots,S_m$, each distributing a physical system to different subsets of $n$ parties $\mc{A}_1,\mc{A}_2,\cdots,\mc{A}_n$ (see Fig.\,\ref{Fig:general_network}). 
Based on the value of his particular input $x_i$, party $\mathcal{A}_i$ performs a local operation collectively on all his shares and generates an output $a_i$.
If the sources are distributing classical variables, then this local operation will be some classical processing; whereas if the sources are distributing quantum states, then the local operation will be a quantum measurement.
In this paper, we assume that both inputs and outputs are binary $(x_i\in\{0,1\},\,a_i\in\{-1,+1\})$. 
The total $n$-partite correlations are characterized by the joint conditional probability distribution $p(a_1,\ldots,a_n\vert x_1,\ldots,x_n)$ over all inputs and outputs. 

Let us identify the structure of classical correlations in this setup.
Suppose that source $S_i$ is distributing classical variable $\lambda_i$, and the joint probability density function (pdf) of all $\lambda_i$ is $\mu(\lambda_1,\cdots,\lambda_m)$.
Let $\Omega_i\subset\{S_1,\cdots S_m\}$ denote the collection of sources that distribute variables to party $\mathcal{A}_i$.
Then every classical correlation has the form
\begin{align}
     &p(a_1,\ldots,a_n\vert x_1,\ldots,x_n)\notag\\
     &=\int d\lambda_1\cdots\int d\lambda_m\,\mu(\lambda_1,\ldots,\lambda_m)\prod_{i=1}^np(a_i\vert x_i,\Omega_i)\label{Eq:Pre:classical-network},
\end{align}
where $p(a_i|x_i,\Omega_i)$ describes the local stochastic processing of party $\mc{A}_i$.
In a quantum model, the shared classical variable $\lambda_i$ is replaced by a multipartite quantum state. 
As shown by Bell \cite{Bell_1964_EPR}, the entanglement in the state can lead to measurement correlations that do not have the form of Eq. \eqref{Eq:Bell-inequality}. 
Such correlations are called nonlocal

In this work, we are interested in how the feasible classical correlations $p(a_1,\ldots,a_n\vert x_1,\ldots,x_n)$ of the parties change as we place restrictions on the interdependency of the source variables $\lambda_i$.
In other words, how do restrictions on the joint pdf $\mu(\lambda_1,\cdots,\lambda_m)$ affect the correlations that can be realized?
We will specifically focus on the strongest restriction of full independence (Sect.\,\ref{Sec:Bell-noSR}), meaning that $\mu(\lambda_1,\cdots,\lambda_m)=\prod_{i=1}^m\mu(\lambda_i)$, and relaxations to this form in variational distance (Sect.\,\ref{Sec:Qnetworks}).

\section{Shared randomness in the minimal Bell scenario}
\label{Sec:Bell-noSR}

%In this section, we study how limitations on shared randomness can be witnessed. We consider a standard Bell setup while the cardinality of the shared random variable $\Lambda$ in the classical model is bounded. Note that, for any single linear Bell inequality, the classical bound can be attained by deterministic strategies. Hence, individual Bell inequalities are unable to distinguish models with different amounts of shared randomness. 

We first consider the simplest case of $m=1$ and $n=2$ (also known as the 2-2-2 scenario).
Suppose the shared random variable $\lambda$ is sampled from some set $\Lambda$.
When $|\Lambda|=1$, there is no genuine randomness shared between the two parties, Alice and Bob, and the input/output correlations factorize as
\begin{align}
\label{Eq:Bell-independent}
    p(a,b|x,y)=p_A(a|x)p_B(b|y).
\end{align}
Despite the output variables being \textit{uncorrelated} in this case, for consistency purposes we still refer to the collection of transition probabilities $p(a,b|x,y)$ as Alice and Bob's \textit{correlations}.
From Eq. \eqref{Eq:Bell-independent}, we see that the classical correlations are nonconvex without shared randomness, and so they cannot be characterized by tight Bell inequalities \cite{Peres-1999a}.
Furthermore, the maximum of any linear expression involving the probabilities $p(a_1,a_2|x_1,x_2)$ can always be obtained classically without shared randomness.
Consequently, a single Bell inequality is always insufficient to capture the limitation of classical correlations satisfying Eq. \eqref{Eq:Bell-independent}.
In the following, we instead investigate the feasible set of correlations under the simultaneous consideration of multiple linear Bell functions, which manifests the nonconvex features of the classical set and allows us to characterize how the feasible region of correlations depends on the presence of shared randomness.

For the (2-2-2) scenario, it is well known that there are only eight nontrivial facets of the local polytope, which can be described by four linear functions of the observed probabilities \cite{Fine_1982_HV}:
\begin{equation}
    \begin{aligned}
        A&=p(00\vert00)-p(01\vert01)-p(10\vert10)-p(00\vert11),\\
        B&=p(00\vert10)-p(01\vert11)-p(10\vert00)-p(00\vert01),\\
        C&=p(00\vert01)-p(01\vert00)-p(10\vert11)-p(00\vert10),\\
        D&=p(00\vert11)-p(01\vert10)-p(10\vert01)-p(00\vert00),
    \end{aligned}
    \label{Eq:4-Bell-func}
\end{equation}
which each range from $-1$ to $0$. 
We will use the combination of these four linear Bell functions as witnesses for shared randomness.

\subsection{Two Bell functions}

To begin with, we consider the simultaneous evaluation of two of the four Bell functions in Eq.\,\eqref{Eq:4-Bell-func}, and without loss of generality, it is sufficient to just focus on three pairs: $(A,B),(A,C),$ and $(A,D)$. This is because other combinations can be transformed into the above cases by relabeling. For example, by flipping the first party's input, the pair $(B,C)$ is mapped to $(A,D)$.

In the case where $\vert\Lambda\vert=1$, define $x_0=p_A(0\vert 0),\,x_1=p_A(0\vert 1),\,y_0=p_B(0\vert 0),\,y_1=p_B(0\vert 1)$.
When combined with normalization constraints, Eq. \eqref{Eq:4-Bell-func} becomes
\begin{equation}
    \begin{aligned}
        A&=x_0y_0-x_0(1-y_1)-(1-x_1)y_0-x_1y_1,\\
        B&=x_1y_0-x_1(1-y_1)-(1-x_0)y_0-x_0y_1,\\
        C&=x_0y_1-x_0(1-y_0)-(1-x_1)y_1-x_1y_0,\\
        D&=x_1y_1-x_1(1-y_0)-(1-x_0)y_1-x_0y_0,
    \end{aligned}
\end{equation}
where $x_0,x_1,y_0,y_1\in[0,1]$. One can verify that for all the three pairs, any classical point $(-p,-q)\in[-1,0]\times[-1,0]$ can be achieved by strategies without shared randomness via explicit constructions. Specifically, $x_0=q,\,x_1=p,\,y_0=0,\,y_1=1$ yields $(A,B)=(-p,-q)$; $x_0=0,\,x_1=1,\,y_0=q,\,y_1=p$ yields $(A,C)=(-p,-q)$; $x_0=1-q,\,x_1=p,\,y_0=0,\,y_1=1$ yields $(A,D)=(-p,-q)$. Therefore, for these pairs, there is no separation between classical models with and without shared randomness.

To manifest the nonconvex structure when $|\Lambda|=1$, we introduce an additional linear function $E$ inspired by the correlation box proposed in \cite{Donohue_2015_idenfy-nonconvexity}:
\begin{align}
    E&=\frac{1}{2}\left[p(00\vert00)+p(11\vert00)+p(00\vert11)+p(11\vert11)\right]\\
    &=\frac{1}{2}\left[x_0y_0+(1-x_0)(1-y_0)+x_1y_1+(1-x_1)(1-y_1)\right],
\end{align}
which ranges from 0 to 1 for all valid probability distributions. Still, by relabeling symmetries, it suffices to consider the pairs $(A,E)$ and $(B,E)$.

\begin{figure}[htbp!]
\centering
\subfigure[]{
  \includegraphics[width=0.45\textwidth]{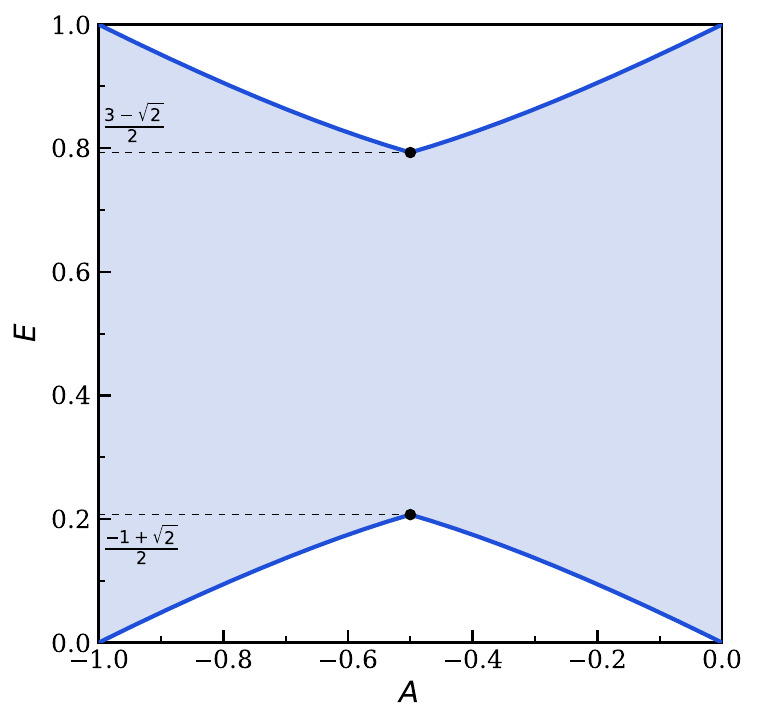}
  \label{Fig:A-E}
}
\hfill
\subfigure[]{
  \includegraphics[width=0.45\textwidth]{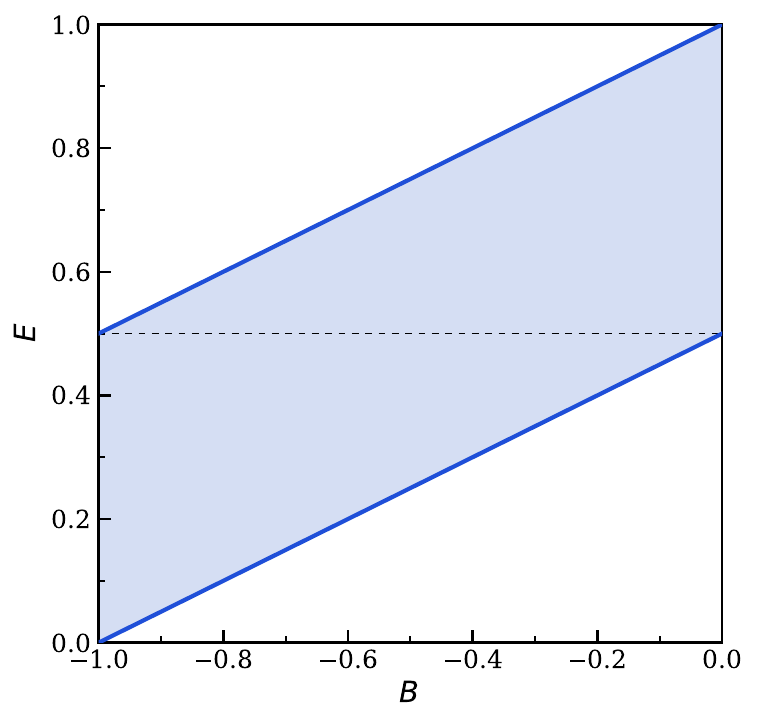}
  \label{Fig:B-E}
}
\caption{
Feasible regions for classical models with $\vert\Lambda\vert=1$. In both panels, the solid curves denote the analytic boundaries of the feasible sets, while the shaded regions represent the attainable values. (a) Joint values of $(A,E)$. (b) Joint values of $(B,E)$.
}
\end{figure}

For the pair $(A,E)$, we numerically compute and plot the feasible region in the $A-E$ plane for strategies without shared randomness (see Fig.\,\ref{Fig:A-E}). The left and right boundaries of the feasible region coincide with those of the unrestricted classical region, while the upper and lower boundaries are nontrivial and can be derived analytically via Karush-Kuhn-Tucker (KKT) conditions. They are given by
\begin{align}
    E_{\max}(A)&=
    \begin{cases}
        1-A-\sqrt{-A}, &A\in[-1,-\frac{1}{2}],\\
        2+A-\sqrt{1+A}, &A\in[-\frac{1}{2},0],
    \end{cases}\\
    E_{\min}(A)&=
    \begin{cases}
        A+\sqrt{-A}, &A\in[-1,-\frac{1}{2}],\\
        -1-A+\sqrt{1+A}, &A\in[-\frac{1}{2},0].
    \end{cases}
\end{align}
Apparently, when the classical dimension is 2, the feasible region already coincides with the unrestricted classical region, since any point in $[-1,0]\times[0,1]$ can be expressed as a convex combination of one point on the left boundary and the other on the right boundary.

For the pair $(B,E)$, the feasible region for strategies without shared randomness is represented in Fig.\,\ref{Fig:B-E}. Any point $(b,e)$ in this region can be achieved by letting $x_0=-b,\,x_1=2e-b-1,\,y_0=0,\,y_1=1$. Note that this region is exactly the same as that of the unrestricted classical model. This is because for any classical behavior,
\begin{align}
    2E-B&=2-p(01\vert00)-p(10\vert11)-p(00\vert10)+p(00\vert01)\notag\\
    &=2+C.
\end{align}
Therefore, $1\leq2E-B\leq2$, and all classical behaviors are bounded in the same region, which means no separation is observed for this pair.

\subsection{Three Bell functions}
\label{Sec:3-Bell}

We now turn to the simultaneous consideration of three of the four linear Bell functions in Eq.\,\eqref{Eq:4-Bell-func}. Up to relabeling, it suffices to consider the triple $(A,B,C)$. The feasible region for strategies without shared randomness is a three-dimensional, nonconvex set. A direct and complete characterization of it is complicated. We therefore focus on its projections by fixing any two coordinates and determining the maximal and minimal achievable values of the third. For example, for given values of $(A,B)$, we compute $C_{\max}(A,B)$ and $C_{\min}(A,B)$, which give out two boundaries of the feasible region. This projection captures the essential structure of the three-dimensional set while remaining tractable.

\begin{figure}[htbp!]
\centering
\subfigure[]{
  \includegraphics[width=0.41\textwidth]{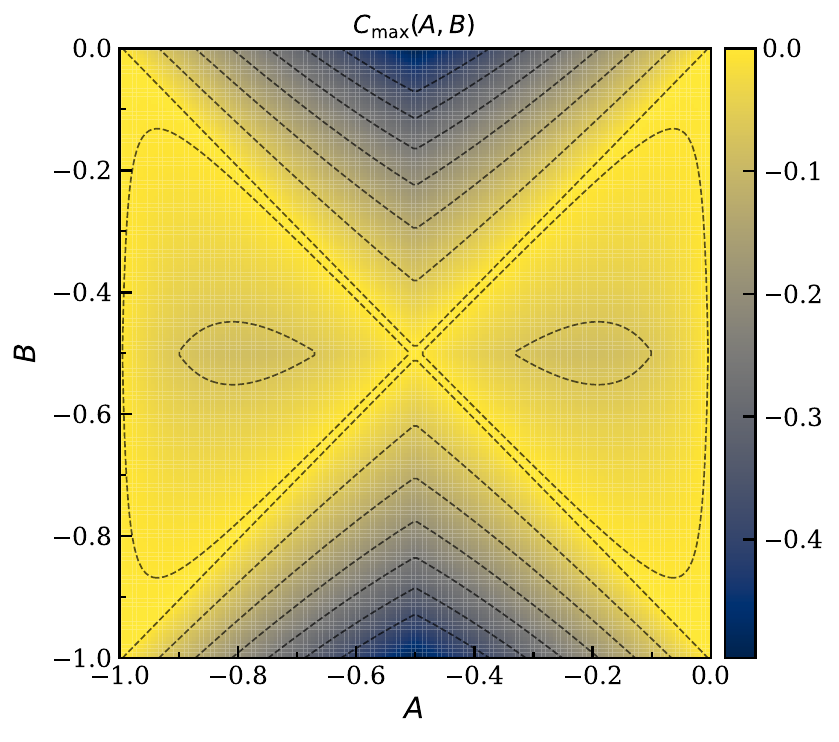}
  \label{Fig:Cmax}
}
\hfill
\subfigure[]{
  \includegraphics[width=0.41\textwidth]{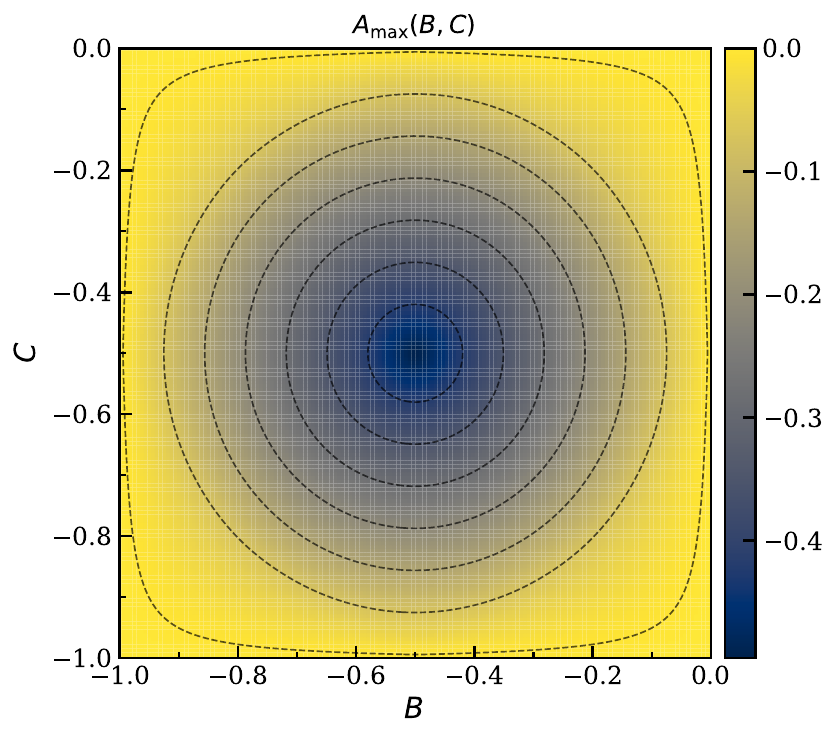}
  \label{Fig:Amax}
}
\hfill
\subfigure[]{
\includegraphics[width=0.41\textwidth]{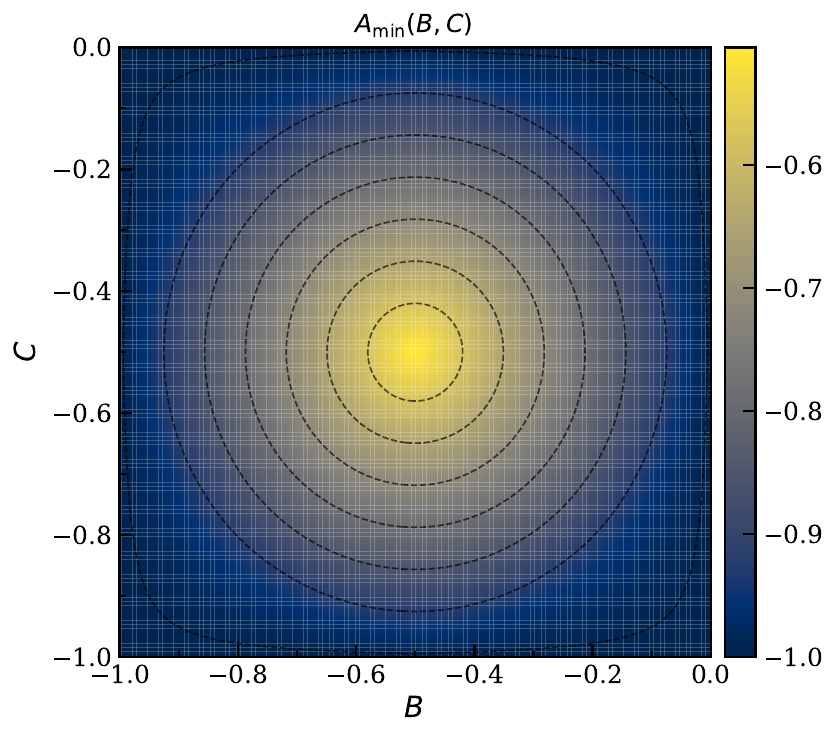}
\label{Fig:Amin}
}
\caption{
Heatmaps of the functions $C_{\max}$, $A_{\max}$, and $A_{\min}$. In all panels, colors represent numerical values of the corresponding function, and the dashed curves indicate contour lines of equal value. (a) $C_{\max}(A,B)$. (b) $A_{\max}(B,C)$. (c) $A_{\min}(B,C)$.
}
\label{Fig:CmaxAmaxAmin}
\end{figure}

The feasible region has six boundaries, corresponding to the maximal and minimal values of each Bell function for fixed values of the other two. Note that by employing symmetries, the problem can be greatly reduced. If a choice of values $(x_0,x_1,y_0,y_1)$ achieves $A=a,B=b,C=c$, then replacing it with
\begin{equation}
    \begin{cases}
        x_0'=x_1,\\
        x_1'=x_0,
    \end{cases}
    \quad
    \begin{cases}
        y_0'=y_0,\\
        y_1'=1-y_1,
    \end{cases}\label{Eq:trans_Cmax-Cmin}
\end{equation}
yields a strategy that achieves $A=a,B=b,C=-1-c$, which means the feasible region is symmetric with respect to the plane $C=-\frac{1}{2}$, and therefore, $C_{\min}=-1-C_{\max}$. Besides, a strategy with
\begin{equation}
    \begin{cases}
        x_0'=y_0,\\
        x_1'=y_1,
    \end{cases}
    \quad
    \begin{cases}
        y_0'=x_0,\\
        y_1'=x_1,
    \end{cases}
\end{equation}
achieves $A=a,B=c,C=b$. Since the point $\left(a,b,C_{\max}(a,b)\right)$ is feasible, $\left(a,C_{\max}(a,b),b\right)$ is also feasible, and hence $B_{\max}\geq C_{\max}$. Conversely, if $\left(a,B_{\max}(a,c),c\right)$ is feasible, then $\left(a,c,B_{\max}(a,c)\right)$ is also feasible. Therefore, $B_{\max}=C_{\max}$. Similarly, $B_{\min}=C_{\min}=-1-C_{\max}$. Consequently, to fully characterize the boundaries of the feasible region, we only need to determine $C_{\max}, A_{\max}$, and $A_{\min}$. We numerically evaluate these functions and present the results in Fig.\,\ref{Fig:CmaxAmaxAmin}.
For $C_{\max}$, by KKT conditions, an analytical form can be derived:
\begin{equation}
\label{Eq:Cmax-3func}
    C_{\max}(A,B)=
    \begin{cases}
        -A\frac{1+A+B}{A+B}, &(A,B)\in\mathcal{R}_1,\\
        (1+A)\frac{1+A+B}{2+A+B}, &(A,B)\in\mathcal{R}_2,\\
        A\frac{A-B}{1-A+B}, &(A,B)\in\mathcal{R}_3,\\
        (1+A)\frac{A-B}{1+A-B}, &(A,B)\in\mathcal{R}_4,
    \end{cases}
\end{equation}
where the ranges are given by
\begin{equation}
    \begin{cases}
        \mathcal{R}_1: A\leq0, B\leq-\frac{1}{2}, A+B\geq-1\ \text{or}\\\quad\;\;\;\ A\leq-\frac{1}{2}, B\leq0, A+B\geq-1,\\
        \mathcal{R}_2: A\geq-\frac{1}{2}, B\geq-1, A+B\leq-1\ \text{or}\\\quad\;\;\;\ A\geq-1, B\geq-\frac{1}{2}, A+B\leq-1,\\
        \mathcal{R}_3: A\leq0, B\geq-\frac{1}{2}, A\geq B\ \text{or}\\\quad\;\;\;\ A\leq-\frac{1}{2}, B\geq-1, A\geq B,\\
        \mathcal{R}_4: A\geq-\frac{1}{2}, B\leq0, A\leq B\ \text{or}\\\quad\;\;\;\ A\geq-1, B\leq-\frac{1}{2}, A\leq B.
    \end{cases}
    \label{Eq:ABC-R}
\end{equation}
For $A_{\max}$ and $A_{\min}$, closed forms are challenging to obtain, which we will discuss in detail in Appendix \ref{App:Amax&Amin}. From the numerical visualization of these boundaries, we observe that any point outside the feasible region but within the classical cube can be expressed as a convex combination of two boundary points. Hence, $\vert\Lambda\vert=2$ suffices to reproduce all classical points in this case.

Equation \eqref{Eq:Cmax-3func} combined with the ranges $\mc{R}_i$ presented in Eq. \eqref{Eq:ABC-R} describe the feasible region for no shared randomness.
Any behavior observed outside this region therefore requires some genuine shared randomness.

\subsection{Four Bell functions}
\label{Sec:4-Bell}

We finally consider the simultaneous evaluation of all four Bell functions $A,B,C,D$. 
Any strategy without shared randomness has to satisfy the equation
\begin{equation}
    (A-B)(A+B+1)=(C-D)(C+D+1),
    \label{Eq:4-Bell-Lambda-1-eq}
\end{equation}
since both sides are equal to $(x_0-x_1)(x_0+x_1-1)(2y_0-1)(2y_1-1)$. Consequently, if the values of any 3 of the 4 are given, take $A,B,C$ as an example, then Eq.\,\eqref{Eq:4-Bell-Lambda-1-eq} becomes a quadratic function of $D$. Hence, the value of $D$ is determined by solving this quadratic. In general, a quadratic may admit two roots, but not both are admissible. For example, if $A=B=C=0$, Eq.\,\eqref{Eq:4-Bell-Lambda-1-eq} has two roots $0,-1$, but one can check that $D=-1$ is not possible for a strategy without shared randomness.

We next extend this observation to consider the attainable values of the quantity
\begin{equation}
    \Delta\equiv A(A+1)+D(D+1)-B(B+1)-C(C+1)
\end{equation}
for different classical dimensions. In particular, we find that
\begin{equation}
    \Delta\in
    \begin{cases}
        \{0\}, &\vert\Lambda\vert=1,\\
        [-\frac{1}{2},\frac{1}{2}], &\vert\Lambda\vert\geq2.\\
    \end{cases}
\end{equation}
The proof is presented in Appendix \ref{App:Proof_bound_dim}. For $\vert\Lambda\vert=1$, $\Delta=0$, and since $D\in[-1,0]$, it follows that $D(D+1)\in[-\frac{1}{4},0]$. Therefore, the quantity
\begin{equation}
    \Theta\equiv B(B+1)+C(C+1)-A(A+1)
\end{equation}
also takes values in the interval $[-\frac{1}{4},0]$. For $\vert\Lambda\vert\geq2$, as we will show in Appendix \ref{App:Proof_bound_dim}, the range of $\Theta$ extends to $[-\frac{1}{2},\frac{1}{4}]$. By the same reasoning, analogous bounds hold for each of the following quantities:
\begin{equation}
    \begin{aligned}
        &B(B+1)+C(C+1)-D(D+1),\\
        &A(A+1)+D(D+1)-B(B+1),\\
        &A(A+1)+D(D+1)-C(C+1).
    \end{aligned}
\end{equation}

\section{Correlated sources in quantum networks}
\label{Sec:Qnetworks}

In this section, we turn to network scenarios involving multiple sources. 
While these sources are typically considered to be independent in standard quantum networks, we study a relaxation of the corresponding classical models by allowing correlations between sources. 
Under this relaxation, standard network inequalities generally fail to constrain classically admissible correlations. 
We therefore quantify correlations between sources and construct nonlinear inequalities based on this characterization. 
The violations of these inequalities imply a no-go for a classical picture even when assisted by some source dependence.

Following the definitions in Sect.\,\ref{Sect:LHV-model}, consider a source-disjoint set of parties $\mathcal{I}=\{i_1,i_2,\ldots,i_k\}$, meaning that $\Omega_{i_j}\cap \Omega_{i_k}=\emptyset$ for all $j,k$. 
For any group of parties in $\mathcal{I}$, the assumption of source independence implies that their marginal correlations admit a description without shared randomness, no matter whether the underlying model is classical or quantum. 
More precisely, the marginal correlations are compatible with a $k$-party LHV model where the global hidden variable $\Lambda$ is trivial; i.e. $|\Lambda|=1$. 
Consequently, we can apply the feasibility constraints derived in Sect.\,\ref{Sec:Bell-noSR} for $\vert\Lambda\vert=1$ on these marginals. 
Any violation of these constraints for any pair of parties in $\mathcal{I}$ will certify the presence of correlations between sources.

To move beyond strict independence, we define the following quantity for any source-disjoint set of parties $\mathcal{I}=\{i_1,i_2,\ldots,i_k\}$,
\begin{align}
    \mathcal{M}_{n,k}=&\inf_{\mu_{i_1},\ldots,\mu_{i_k}}\int d\Omega_{i_1}\cdots\int d\Omega_{i_k}\notag\\
    &\times\left\vert\mu(\Omega_{i_1},\ldots,\Omega_{i_k})-\prod_{j=1}^k\mu_{i_j}(\Omega_{i_j})\right\vert.
\end{align}
Here, $d\Omega_{i_j}=\prod_{\lambda\in\Omega_{i_j}}d\lambda$, $\mu(\Omega_{i_j})$ is the joint probability distribution of all classical variables in the set $\Omega_{i_j}$, and the infimum is taken over all product distributions of the form $\prod_{j=1}^k\mu_{i_j}(\Omega_{i_j})$. 
It is worth noting that $\mathcal{M}_{n,k}=0$ implies that the sets of source $\{\Omega_{i_1},\cdots,\Omega_{i_k}\}$  are independent, but sources within the same set may still be correlated. 
We further define two quantities $I_{n,k}(\{x_i,i\in\overline{\mathcal{I}}\})$ and $J_{n,k}(\{x_i,i\in\overline{\mathcal{I}}\})$ as
\begin{equation}
    \begin{aligned}
        I_{n,k}(\{x_i,i\in\overline{\mathcal{I}}\})&=\frac{1}{2^k}\!\sum_{x_j,\,j\in\mathcal{I}}\!\langle A^1_{x_1}\cdots A^n_{x_n}\rangle,\\
        J_{n,k}(\{x_i,i\in\overline{\mathcal{I}}\})&=\frac{1}{2^k}\!\sum_{x_j,\,j\in\mathcal{I}}\!{(-1)}^{\sum_{l\in\mathcal{I}}x_l}\langle A^1_{x_1}\cdots A^n_{x_n}\rangle.
    \end{aligned}
    \label{Eq:Def-I-J}
\end{equation}
Note that the summations above are taken only over the inputs $x_j$ with $j\in\mc{I}$; the inputs $x_i$ are fixed for $i\in\overline{\mc{I}}$.
\begin{proposition}
\label{Prop:ineq}
For any classical model whose source dependence is characterized by $\mathcal{M}_{n,k}$, the generated correlations must satisfy the following inequality
\begin{equation}
    {\left(\vert I_{n,k}(\{x_i\})\vert-\mathcal{M}_{n,k}\right)}^{\frac{1}{k}}+{\left(\vert J_{n,k}(\{x_i'\})\vert-\mathcal{M}_{n,k}\right)}^{\frac{1}{k}}\leq1,
    \label{Eq:I-J-M}
\end{equation}
for all choices of inputs on the other $n-k$ in $\overline{\mc{I}}$. 
\end{proposition}
\noindent Note that in Eq.\,\eqref{Eq:I-J-M}, the values of $x_i$ and $x_i'$ may be chosen differently. The proof of Proposition \ref{Prop:ineq} is presented in Appendix \ref{Sec:app:Pf:I-J-M}.

\begin{figure}[b]
\centering
\begin{tikzpicture}[
  font=\large,
  >=Latex,
  line cap=round,
  line join=round,
  % -------------------------
  % Styles
  % -------------------------
  party/.style={
    circle, draw=black, line width=0.9pt,
    minimum size=12mm,
    fill=gray!15
  },
  inarrow/.style={
    -{Latex[length=2.5mm,width=2mm]},
    line width=1.0pt
  },
  outarrow/.style={
    -{Latex[length=2.5mm,width=2mm]},
    line width=1.0pt
  },
  hiddenbi/.style={
    dashed,
    line width=1.0pt,
    <->,
    {Latex[length=2.5mm,width=2mm]}-{Latex[length=2.5mm,width=2mm]}
  },
  dots/.style={font=\Large}
]

% -------------------------
% Geometry control
% -------------------------
\def\R{3.0} % radius of star

% -------------------------
% Central party B
% -------------------------
\node[party] (B) at (0,0) {$\mathcal{B}$};

% Input y and output b for B
\node (y) at (0,1.35) {$y$};
\draw[inarrow] (y) -- (B.north);

\node (b) at (0,-1.35) {$b$};
\draw[outarrow] (B.south) -- (b);

% -------------------------
% Outer parties (explicit + ellipsis)
% -------------------------
\node[party] (A1) at ({\R*cos(135)},{\R*sin(135)}) {$\mathcal{A}_{1}$};
\node[party] (A2) at ({\R*cos(45)},{\R*sin(45)}) {$\mathcal{A}_{2}$};
\node[party] (Anm) at ({\R*cos(-45)},{\R*sin(-45)}) {$\mathcal{A}_{N-1}$};
\node[party] (An) at ({\R*cos(-135)},{\R*sin(-135)}) {$\mathcal{A}_{N}$};

% Ellipses indicating many parties/sources
\node[dots] at ({\R*cos(0)},{\R*sin(0)}) {$\boldsymbol{\cdots}$};

% -------------------------
% Bidirectional hidden links with Lambda_i labels
% -------------------------
\draw[hiddenbi] (B) -- (A1)
  node[midway, above, sloped] {$S_{1}$};

\draw[hiddenbi] (B) -- (A2)
  node[midway, above, sloped] {$S_{2}$};

\draw[hiddenbi] (B) -- (Anm)
  node[midway, above, sloped] {$S_{N-1}$};

\draw[hiddenbi] (B) -- (An)
  node[midway, above, sloped] {$S_{N}$};

% -------------------------
% Inputs x_i and outputs a_i for outer parties
% -------------------------
% A1
\node (x1) at ($(A1)+(-0.9,1.0)$) {$x_{1}$};
\draw[inarrow] (x1) -- ($(A1.north west)+(0.35,0.1)$);
\node (a1) at ($(A1)+(0,-1.35)$) {$a_{1}$};
\draw[outarrow] (A1.south) -- (a1);

% A2
\node (x2) at ($(A2)+(0.9,1.0)$) {$x_{2}$};
\draw[inarrow] (x2) -- ($(A2.north east)+(-0.35,0.1)$);
\node (a2) at ($(A2)+(0,-1.35)$) {$a_{2}$};
\draw[outarrow] (A2.south) -- (a2);

% A_{n-1}
\node (xnm) at ($(Anm)+(0.9,1.0)$) {$x_{N-1}$};
\draw[inarrow] (xnm) -- ($(Anm.north east)+(-0.35,0.1)$);
\node (anm) at ($(Anm)+(0,-1.35)$) {$a_{N-1}$};
\draw[outarrow] (Anm.south) -- (anm);

% A_n
\node (xn) at ($(An)+(-0.9,1.0)$) {$x_{N}$};
\draw[inarrow] (xn) -- ($(An.north west)+(0.35,0.1)$);
\node (an) at ($(An)+(0,-1.35)$) {$a_{N}$};
\draw[outarrow] (An.south) -- (an);

\end{tikzpicture}
\caption{A star network with $N$ branches and a central node. Each branch $\mathcal{A}_i$ and the central node $\mathcal{B}$ share a source $S_i$.}
\label{Fig:starnetwork}
\end{figure}
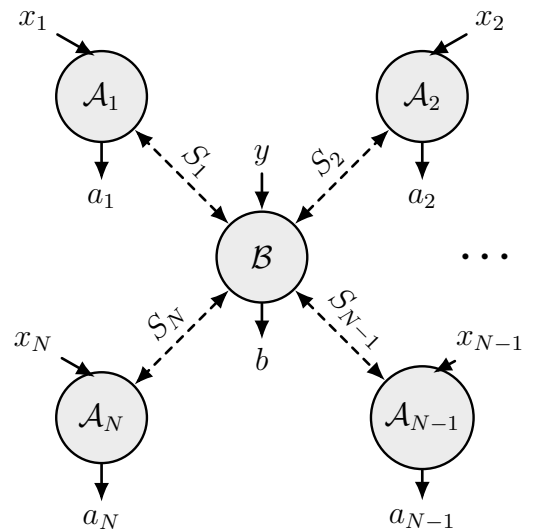

As an example, we consider a star-shaped network shown in Fig.\,\ref{Fig:starnetwork}, where the central party is labeled by $\mathcal{B}$, with corresponding observables $B_0$ and $B_1$. For this star topology, there are $N$ source-disjoint parties $\mathcal{A}_1,\mathcal{A}_2,\ldots,\mathcal{A}_N$, with $\Omega_i=\{S_i\}, 1\leq i\leq N$. 
Then Proposition \ref{Prop:ineq} says that
\begin{equation}
    {\left(\vert I_N\vert-\mathcal{M}_N\right)}^{\frac{1}{N}}+{\left(\vert J_N\vert-\mathcal{M}_N\right)}^{\frac{1}{N}}\leq 1,
\end{equation}
where
\begin{equation}
    \begin{aligned}
        I_N&=\frac{1}{2^N}\sum_{x_1\cdots x_N}\langle A^1_{x_1}\cdots A^N_{x_N}B_0\rangle,\\
        J_N&=\frac{1}{2^N}\sum_{x_1\cdots x_N}{(-1)}^{\sum_{i=1}^N x_i}\langle A_{x_1}^1\cdots A_{x_N}^NB_1\rangle.
    \end{aligned}
\end{equation}
and
\[ \mathcal{M}_N=\inf_{\{\mu_i\}}\int d\lambda_1\cdots\int d\lambda_N\vert\mu(\lambda_1,\ldots,\lambda_N)-\prod_{i=1}^N\mu_i(\lambda_i)\vert.\]

Next, we discuss the quantum violation of this inequality without any relaxation on the source independence. Assume that each source $S_i$ emits two particles in the two-qubit Werner state $\rho_w(v_i)=v_i\ketbra{\Phi^+}{\Phi^+}+(1-v_i)I_4/4$, where $v_i\in[0,1]$, $\ket{\Phi^+}=(\ket{00}+\ket{11})/\sqrt{2}$, and $I_4$ is the four-dimensional identity operator. Define $V=\prod_{i=1}^Nv_i$. Consider the measurement strategy given by $A_0^i=(\sigma_z+\sigma_x)/\sqrt{2}$ and $A_1^i=(\sigma_z-\sigma_x)/\sqrt{2}$ for all $1\leq i\leq N$, $B_0=\sigma_z^{\otimes N}$ and $B_1=\sigma_x^{\otimes N}$. With these choices, the nonclassicality can be manifested for any $V>{(\frac{1}{\sqrt{2}})}^N+{\sqrt{2}}^N\mathcal{M}$.

As another application of Proposition \ref{Prop:ineq}, we return to the bipartite setting and derive a Bell inequality under limited shared randomness. 
Suppose that the two parties, Alice and Bob, share a common source of randomness $S$ quantified by its entropy $H(S)$. 
Note that any classical correlation in this Bell setting can be simulated by a network where Alice and Bob receive sources $S_1$ and $S_2$, respectively, with $I(S_1:S_2)=H(S)$. By applying Eq.\,\eqref{Eq:I-J-M} to this network, we obtain
\begin{align}
    1\geq &\sqrt{\left\vert \left\langle\frac{A_0+A_1}{2}\frac{B_0+B_1}{2}\right\rangle \right\vert-\mathcal{M}}\,\notag\\
    &+\,\sqrt{\left\vert \left\langle\frac{A_0-A_1}{2}\frac{B_0-B_1}{2}\right\rangle \right\vert-\mathcal{M}},
\end{align}
where $A_0,A_1$ and $B_0,B_1$ are observables for Alice and Bob. 
Additionally, we have
\begin{align}
    \mathcal{M}&=\inf_{\mu_1,\mu_2}\int d\lambda_1\int d\lambda_2\vert\mu(\lambda_1,\lambda_2)-\mu_1(\lambda_1)\mu_2(\lambda_2)\vert\notag\\
    &\leq\int d\lambda_1\int d\lambda_2\vert\mu(\lambda_1,\lambda_2)-\mu(\lambda_1)\mu(\lambda_2)\vert\notag\\
    &\leq\sqrt{2\ln2\,I(S_1:S_2)},
\end{align}
where the last line comes from Pinsker's inequality \cite{Fedotov_2003_Pinsker}. Therefore, any classical correlation in a bipartite Bell scenario with shared randomness bounded by $H(S)$ satisfies the following inequality:
\begin{align}
    1\geq&\sqrt{\left\vert \left\langle\frac{A_0+A_1}{2}\frac{B_0+B_1}{2}\right\rangle \right\vert-\sqrt{2\ln2\,H(S)}}\,\notag\\
    &+\,\sqrt{\left\vert \left\langle\frac{A_0-A_1}{2}\frac{B_0-B_1}{2}\right\rangle \right\vert-\sqrt{2\ln2\,H(S)}}.
\end{align}
This provides a new entropic relation for the effect of limited shared randomness in the standard bipartite Bell scenario.

\section{Discussion}

In this work, we have studied local models of Bell nonlocality under different types of restrictions and relaxations placed on the classical resources. 
We first considered the bipartite Bell scenarios with bounded classical dimension and characterized the feasible region with no shared randomness through the simultaneous evaluation of multiple linear Bell functions. 
Our work provides nonlinear inequalities that can witness shared randomness by considering as few as three different linear Bell functions. 
We then turned to network scenarios and studied relaxations of the classical models by allowing correlations between sources. 
We showed that feasibility constraints derived in the bipartite setting can be used to certify source dependence, and we constructed nonlinear inequalities that separate classical models with correlated sources from correlations attainable in standard source-independent networks. 
As an application to the bipartite case, we obtained an entropic Bell inequality under limited shared randomness.

Beyond the specific scenarios considered here, the simultaneous consideration of multiple linear Bell functions introduced in this work can be applied in a much broader context. In this paper, we restricted our attention mainly to the CHSH facet inequality. 
It would be interesting to explore other choices of linear Bell functions and investigate how different combinations of such functions give rise to richer feasibility structures and enhanced witnessing power. 
Additionally, while we focused on the simplest bipartite Bell scenario, a further generalization direction is to consider witnessing shared randomness in an $n$-party Bell scenario with a single source.

From a methodological perspective, it would also be interesting to understand the relationship between the number of linear Bell functions considered and the largest classical dimension that can be witnessed. 
For a fixed number $m$ of linear Bell functions, the classical feasible region is a convex subset of $\mathbb{R}^m$. It therefore follows from Carath\'eodory's theorem \cite{Caratheodory_1907} that every point in this region can be expressed as a convex combination of at most $m+1$ points. Consequently, classical models of dimension $m+1$ already suffice to span the entire classical feasible region. This observation leads to two related future directions. The first one is whether this bound $m+1$ is tight, and the second one is to characterize the relationship among those $m$ Bell functions that optimally approach or attain this bound, which may allow for a deeper understanding of the relationship between different Bell inequalities.

\begin{acknowledgments}
This work was funded by a Research Experience for Undergraduates (REU) supplement to  National Science Foundation (NSF) grant No. 2112890, as well as NSF grant No. 2426154.

\end{acknowledgments}

\section{Appendixes}

\appendix
\section{Explicit constructions for boundary points}

In this appendix, we provide explicit constructions that achieve the boundary values discussed in Sect.\,\ref{Sec:Bell-noSR} for reference. The first one is the boundary represented in Fig.\,\ref{Fig:A-E}. The upper boundary $E_{\max}(A)$ can be achieved by
\begin{equation}
    \begin{cases}
        \begin{cases}
            x_0=y_0=0,\\
            x_1=y_1=\sqrt{-A},
        \end{cases}
        & A\in[-1,-\frac{1}{2}],\\
        \begin{cases}
            x_0=y_0=\sqrt{1+A},\\
            x_1=y_1=1,
        \end{cases}
        & A\in[-\frac{1}{2},0],
    \end{cases}
\end{equation}
while the lower boundary $E_{\min}(A)$ can be achieved by
\begin{equation}
    \begin{cases}
        \begin{cases}
            x_0=1-\sqrt{-A},\\
            x_1=0,
        \end{cases}\
        \begin{cases}
            y_0=\sqrt{-A},\\
            y_1=1,
        \end{cases}
        & A\in[-1,-\frac{1}{2}],\\
        \begin{cases}
            x_0=0,\\
            x_1=\sqrt{1+A},
        \end{cases}\
        \begin{cases}
            y_0=1,\\
            y_1=1-\sqrt{1+A},
        \end{cases}
        & A\in[-\frac{1}{2},0].
    \end{cases}
\end{equation}
The left boundary $A=-1$ can be realized by
\begin{equation}
    \begin{cases}
        \begin{cases}
            x_0=0,\\
            x_1=0,
        \end{cases}\
        \begin{cases}
            y_0=1,\\
            y_1=1-2E,
        \end{cases}
        & E\in[0,\frac{1}{2}],\\
        \begin{cases}
            x_0=2(1-E),\\
            x_1=1,
        \end{cases}\
        \begin{cases}
            y_0=0,\\
            y_1=1,
        \end{cases}
        & E\in[\frac{1}{2},1],
    \end{cases}
\end{equation}
while the right boundary $A=0$ can be realized by
\begin{equation}
    \begin{cases}
        \begin{cases}
            x_0=2E,\\
            x_1=1,
        \end{cases}\
        \begin{cases}
            y_0=1,\\
            y_1=0,
        \end{cases}
        & E\in[0,\frac{1}{2}],\\
        \begin{cases}
            x_0=0,\\
            x_1=0,
        \end{cases}\
        \begin{cases}
            y_0=0,\\
            y_1=2(1-E),
        \end{cases}
        & E\in[\frac{1}{2},1].
    \end{cases}
\end{equation}

We next consider the boundary $C_{\max}(A,B)$ illustrated in Fig.\,\ref{Fig:Cmax}, which can be achieved by
\begin{equation}
    \begin{cases}
        \begin{cases}
            x_0=0,\\
            x_1=-A-B,
        \end{cases}\
        \begin{cases}
            y_0=0,\\
            y_1=\frac{A}{A+B},
        \end{cases}
        &(A,B)\in\mathcal{R}_1,\\
        \begin{cases}
            x_0=2+A+B,\\
            x_1=0,
        \end{cases}\
        \begin{cases}
            y_0=1,\\
            y_1=\frac{1+A}{2+A+B},
        \end{cases}
        &(A,B)\in\mathcal{R}_2,\\
        \begin{cases}
            x_0=0,\\
            x_1=A-B,
        \end{cases}\
        \begin{cases}
            y_0=\frac{-A}{1-A+B},\\
            y_1=0,
        \end{cases}
        &(A,B)\in\mathcal{R}_3,\\
        \begin{cases}
            x_0=B-A,\\
            x_1=0,
        \end{cases}\
        \begin{cases}
            y_0=\frac{-B}{1+A-B},\\
            y_1=0,
        \end{cases}
        &(A,B)\in\mathcal{R}_4,
    \end{cases}
    \label{Eq:app:explicit_constructions_Cmax}
\end{equation}
where $\mathcal{R}_1,\mathcal{R}_2,\mathcal{R}_3,\mathcal{R}_4$ take their definitions from \eqref{Eq:ABC-R}.

\section{Analysis of $A_{\max}$ and $A_{\min}$}
\label{App:Amax&Amin}

In this appendix, we analyze the boundary functions $A_{\max}$ and $A_{\min}$ in Sect.\,\ref{Sec:3-Bell}. While the two functions can be characterized numerically, obtaining analytical expressions is more challenging than the case of $C_{\max}$. We shall briefly explain where this difficulty comes from.

We first consider the derivation of $C_{\max}$. We introduce KKT multipliers $\lambda_1,\lambda_2$, and $\mu_i$, and define the following Lagrangian:
\begin{equation}
    \begin{aligned}
        L_1=& x_0y_1-x_0(1-y_0)-(1-x_1)y_1-x_1y_0\\
        &+\lambda_1\left[x_0y_0-x_0(1-y_1)-(1-x_1)y_0-x_1y_1-A\right]\\
        &+\lambda_2\left[x_1y_0-x_1(1-y_1)-(1-x_0)y_0-x_0y_1-B\right]\\
        &+\mu_1x_0+\mu_2(1-x_0)+\mu_3x_1+\mu_4(1-x_1)\\
        &+\mu_5y_0+\mu_6(1-y_0)+\mu_7y_1+\mu_8(1-y_1).
    \end{aligned}
\end{equation}
The local maxima must satisfy the KKT conditions:
\begin{align}
    &\begin{cases}
        (\lambda_1+\lambda_2+1)x_0+(\lambda_1+\lambda_2-1)x_1=\lambda_1+\lambda_2-\mu_5+\mu_6,\\
        (\lambda_1-\lambda_2+1)x_0+(\lambda_2-\lambda_1+1)x_1=1-\mu_7+\mu_8,
    \end{cases}\label{Eq:app:Cmax_KKT_grad-x}\\
    &\begin{cases}
        (\lambda_1+\lambda_2+1)y_0+(\lambda_1-\lambda_2+1)y_1=\lambda_1+1-\mu_1+\mu_2,\\
        (\lambda_1+\lambda_2+1)y_0+(\lambda_2-\lambda_1-1)y_1=\lambda_2-\mu_3+\mu_4,
    \end{cases}\label{Eq:app:Cmax_KKT_grad-y}\\
    &\begin{cases}
        x_0y_0-x_0(1-y_1)-(1-x_1)y_0-x_1y_1=A,\\
        x_1y_0-x_1(1-y_1)-(1-x_0)y_0-x_0y_1=B,
    \end{cases}\label{Eq:app:Cmax_KKT_ABcons}\\
    &\ 0\leq x_0,x_1,y_0,y_1\leq1,\quad\mu_i\geq0,\notag\\
    &\begin{cases}
        \mu_1x_0=\mu_2(1-x_0)=\mu_3x_1=\mu_4(1-x_1)=\\
        \mu_5y_0=\mu_6(1-y_0)=\mu_7y_1=\mu_8(1-y_1)=0.
    \end{cases}\label{Eq:app:Cmax_KKT_mu-x}
\end{align}
Label the determinants of Eq.\,\eqref{Eq:app:Cmax_KKT_grad-x} and \eqref{Eq:app:Cmax_KKT_grad-y} as $M_x,M_y$.
\begin{equation}
    \begin{aligned}
        M_x&=2(1+\lambda_2^2-\lambda_1^2),\\
        M_y&=2(\lambda_2+\lambda_1+1)(\lambda_2-\lambda_1-1).
    \end{aligned}
\end{equation}
We first consider the constraints in Eq.\,\eqref{Eq:app:Cmax_KKT_mu-x}. If all multipliers $\mu_i$ vanish, it is equivalent to unconstrained Lagrangian optimization in the interior. In this case, if both $M_x$ and $M_y$ are nonzero, $x_0=x_1=y_0=y_1=\frac{1}{2}$, $L_1=-\frac{1}{2}$. If only one of them is zero, for example, $M_x=0$, then $y_0=y_1=\frac{1}{2}$, and $\lambda_1,\lambda_2,x_0,x_1$ remain undetermined. Since there are four constraints: $M_x=0$, the single independent equation from Eq.\,\eqref{Eq:app:Cmax_KKT_grad-x}, and the two equations in Eq.\,\eqref{Eq:app:Cmax_KKT_ABcons}, $x_0$ and $x_1$ can be solved as functions of $A$ and $B$, which determines the value of $L_1$. If both $M_x$ and $M_y$ are zero, the following six constraints: $M_x=M_y=0$, one independent equation from Eq.\,\eqref{Eq:app:Cmax_KKT_grad-x}, one independent equation from Eq.\,\eqref{Eq:app:Cmax_KKT_grad-y}, and the two equations in Eq.\,\eqref{Eq:app:Cmax_KKT_ABcons} allow one to solve for $x_0,x_1,y_0,y_1,\lambda_1,\lambda_2$, and therefore determine the value of $L_1$. On the other hand, if some of the $\mu_i$ are nonzero, Eq.\,\eqref{Eq:app:Cmax_KKT_ABcons} implies that the corresponding variables are fixed at the boundary of the feasible domain. The total number of variables remains unchanged, because introducing a multiplier $\mu_i$ simultaneously fixes one variable $x$ or $y$. Therefore, the system remains solvable. Finally, by maximizing over all local maxima, we obtain the piecewise function $C_{\max}$.

We next consider the derivation of $A_{\max}$ as an example. Similarly, by introducing KKT multipliers $\lambda_1,\lambda_2$, and $\mu_i$, we define the following Lagrangian:
\begin{equation}
    \begin{aligned}
        L_2=& x_0y_0-x_0(1-y_1)-(1-x_1)y_0-x_1y_1\\
        &+\lambda_1\left[x_1y_0-x_1(1-y_1)-(1-x_0)y_0-x_0y_1-B\right]\\
        &+\lambda_2\left[x_0y_1-x_0(1-y_0)-(1-x_1)y_1-x_1y_0-C\right]\\
        &+\mu_1x_0+\mu_2(1-x_0)+\mu_3x_1+\mu_4(1-x_1)\\
        &+\mu_5y_0+\mu_6(1-y_0)+\mu_7y_1+\mu_8(1-y_1).
    \end{aligned}
\end{equation}
The local maxima must satisfy the following KKT conditions:
\begin{align}
    &\begin{cases}
        (\lambda_1+\lambda_2+1)x_0+(\lambda_1-\lambda_2+1)x_1=1+\lambda_1-\mu_5+\mu_6,\\
        (\lambda_2-\lambda_1+1)x_0+(\lambda_1+\lambda_2-1)x_1=\lambda_2-\mu_7+\mu_8,
    \end{cases}\label{Eq:app:Amax_KKT_grad-x}\\
    &\begin{cases}
        (\lambda_1+\lambda_2+1)y_0+(\lambda_2-\lambda_1+1)y_1=\lambda_2+1-\mu_1+\mu_2,\\
        (\lambda_1-\lambda_2+1)y_0+(\lambda_1+\lambda_2-1)y_1=\lambda_1-\mu_3+\mu_4,
    \end{cases}\label{Eq:app:Amax_KKT_grad-y}\\
    &\begin{cases}
        x_1y_0-x_1(1-y_1)-(1-x_0)y_0-x_0y_1=B,\\
        x_0y_1-x_0(1-y_0)-(1-x_1)y_1-x_1y_0=C,
    \end{cases}\label{Eq:app:Amax_KKT_ABcons}\\
    &\ 0\leq x_0,x_1,y_0,y_1\leq1,\quad\mu_i\geq0,\notag\\
    &\begin{cases}
        \mu_1x_0=\mu_2(1-x_0)=\mu_3x_1=\mu_4(1-x_1)=\\
        \mu_5y_0=\mu_6(1-y_0)=\mu_7y_1=\mu_8(1-y_1)=0.
    \end{cases}\label{Eq:app:Amax_KKT_mu-x}
\end{align}
A major difference from the case of $C_{\max}$ is that the determinants of Eq.\,\eqref{Eq:app:Amax_KKT_grad-x} and \eqref{Eq:app:Amax_KKT_grad-y} are the same, labeled by
\begin{equation}
    N=2(\lambda_1^2+\lambda_2^2-1).
\end{equation}
We first consider the case when all multipliers $\mu_i$ vanish. If $N$ is nonzero, $x_0=x_1=y_0=y_1=\frac{1}{2}$, $L_2=-\frac{1}{2}$. If $N=0$, there are six parameters $x_0,x_1,y_0,y_1,\lambda_1,\lambda_2$, while there are only five constraints in total: $N=0$, one independent equation from Eq.\,\eqref{Eq:app:Amax_KKT_grad-x}, one independent equation from Eq.\,\eqref{Eq:app:Amax_KKT_grad-y}, and the two equations in Eq.\,\eqref{Eq:app:Amax_KKT_ABcons}, which means one free parameter is subject to optimization. This optimization is too challenging to be tackled analytically because the constraints are quadratic functions of parameters. The same problem arises when seeking the local maxima on the boundary.

In the last part of this appendix, we comment on the relationship between $A_{\max}$ and $A_{\min}$. Numerical results suggest that $A_{\max}=-1-A_{\min}$. However, we fail to find a simple transformation like Eq.\,\eqref{Eq:trans_Cmax-Cmin} to prove this symmetry. For example, on the boundary $C_{\max}$, the ranges $\mathcal{R}_1$ and $\mathcal{R}_4$ are mapped into each other under the transformation $A\rightarrow-1-A$, and similarly for $\mathcal{R}_2$ and $\mathcal{R}_3$. However, when examining the explicit constructions that achieve the boundary $C_{\max}$ in Eq.\,\eqref{Eq:app:explicit_constructions_Cmax}, we do not find a simple transformation acting on the variables $(x_0,x_1,y_0,y_1)$ that maps a strategy achieving a point in $\mathcal{R}_1$ to a strategy achieving the corresponding symmetric point in $\mathcal{R}_4$.

\section{Proof of bounds in Sect.\,\ref{Sec:4-Bell}}
\label{App:Proof_bound_dim}

In this appendix, we provide proofs of the attainable values of $\Delta$ and $\Theta$ for different classical dimensions. First, for $\Delta$, when $\vert\Lambda\vert=1$, it can only take the value 0, which follows directly from Eq.\,\eqref{Eq:4-Bell-Lambda-1-eq}. Note that for all unrestricted classical strategies, feasible values are bounded in $[-\frac{1}{2},\frac{1}{2}]$, because each term, for example, $A(A+1)$, is bounded in $[-\frac{1}{4},0]$. Therefore, we only need to show that each point in $[-\frac{1}{2},\frac{1}{2}]$ can be achieved by $\vert\Lambda\vert=2$.

In the case of $\vert\Lambda\vert=2$, $A=pA_1+(1-p)A_2$, where $p\in[0,1]$ and $A_1,A_2$ are values achieved by two strategies without shared randomness, and similarly for $B,C$ and $D$. Therefore,
\begin{align}
    \Delta=-p(1-p)&[{(A_1-A_2)}^2+{(D_1-D_2)}^2\notag\\
    &-{(B_1-B_2)}^2-{(C_1-C_2)}^2].
\end{align}
By choosing $x_0^1=y_0^1=0,\,x_1^1=y_1^1=1$, we have $A_1=D_1=-1,B_1=C_1=0$, and by choosing $x_0^2=x_1^2=y_0^2=y_1^2=0$, we have $A_2=B_2=C_2=D_2=0$. In this case, $\Delta=-2p(1-p)$, and all values from $[-\frac{1}{2},0]$ are achievable. Besides, by choosing $x_0^1=y_1^1=0,\,x_1^1=y_0^1=1$, $A_1=D_1=0,B_1=C_1=-1$; $x_0^2=x_1^2=y_0^2=y_1^2=0$, $A_2=B_2=C_2=D_2=0$. In this case, $\Delta=2p(1-p)$, and all values from $[0,\frac{1}{2}]$ are achievable, which concludes the proof.

We next consider the attainable values for $\Theta$ when $\vert\Lambda\vert\geq2$. First, for all unrestricted classical strategies, feasible values are bounded in $[-\frac{1}{2},\frac{1}{4}]$ since each term takes values in $[-\frac{1}{4},0]$. We then move on to show that $\vert\Lambda\vert=2$ is able to saturate. When $\vert\Lambda\vert=2$,
\begin{align}
    \Theta=&D(D+1)+p(1-p)[{(A_1-A_2)}^2+{(D_1-D_2)}^2\notag\\
    &-{(B_1-B_2)}^2-{(C_1-C_2)}^2].
\end{align}
By choosing $x_0^1=y_1^1=0,\,x_1^1=y_0^1=1$, $A_1=D_1=0,B_1=C_1=-1$; $x_0^2=x_1^2=y_0^2=y_1^2=0$, $A_2=B_2=C_2=D_2=0$, $\Theta=-2p(1-p)\in[-\frac{1}{2},0]$. By choosing $x_0^1=x_1^1=y_1^1=0,\,y_0^1=1$, $A_1=B_1=-1,C_1=D_1=0$; $x_0^2=y_0^2=y_1^2=0,\,x_1^2=1$, $A_2=C_2=0,B_2=D_2=-1$, $\Theta=p(1-p)\in[0,\frac{1}{4}]$.

\section{Proof of Eq.\,\eqref{Eq:I-J-M}}
\label{Sec:app:Pf:I-J-M}

From Eq.\,\eqref{Eq:Pre:classical-network}, the expectation values in a classical model can be written as
    \begin{align}
        &\langle A_{x_1}^1A_{x_2}^2\cdots A_{x_n}^n\rangle \notag\\
        =&\sum_{a_1\cdots a_n}\left(\prod_{i=1}^na_i\right)p(a_1,a_2,\ldots,a_n\vert x_1,x_2,\ldots,x_n)\notag\\
        =&\int d\lambda_1\int d\lambda_2\cdots\int d\lambda_m\mu(\lambda_1,\lambda_2,\ldots,\lambda_m)\notag\\
        &\quad\quad\quad\prod_{i=1}^n\left(\sum_{a_i}a_ip(a_i\vert x_i,\Omega_i)\right).
    \end{align}
Denote
\begin{equation}
    \langle A^i_{x_i}(\Omega_i)\rangle=\sum_{a_i}a_ip(a_i\vert x_i,\Omega_i),
\end{equation}
we have $\vert \langle A^i_{x_i}(\Omega_i)\rangle \vert\leq1$. Hence, we can obtain
    \begin{align}
        \vert I_{n,k}\vert&\leq\frac{1}{2^k}\int d\lambda_1\int d\lambda_2\cdots\int d\lambda_m\mu(\lambda_1,\lambda_2,\ldots,\lambda_m)\notag\\
        &\quad\quad\quad\prod_{i\in\mathcal{I}}\vert\langle A^i_{0}(\Omega_i)\rangle+\langle A^i_{1}(\Omega_i)\rangle\vert\prod_{j\in\overline{\mathcal{I}}}\vert\langle A^j_{x_j}(\Omega_j)\rangle\vert\notag\\
        &\leq\int d\lambda_1\int d\lambda_2\cdots\int d\lambda_m\mu(\lambda_1,\lambda_2,\ldots,\lambda_m)\notag\\
        &\quad\quad\prod_{i\in\mathcal{I}}\left\vert \frac{\langle A^i_{0}(\Omega_i)\rangle+\langle A^i_{1}(\Omega_i)\rangle}{2} \right\vert.
    \end{align}
By setting $\langle\Delta_{\pm}A^i(\Omega_i)\rangle=(\langle A^i_{0}(\Omega_i)\rangle\pm\langle A^i_{1}(\Omega_i)\rangle)/2$, the above inequality yields to
\begin{align}
        &\vert I_{n,k}\vert\notag\\
        \leq&\int d\lambda_1\int d\lambda_2\cdots\int d\lambda_m\mu(\lambda_1,\lambda_2,\ldots,\lambda_m)\prod_{i\in\mathcal{I}}\vert \langle \Delta_+A^i(\Omega_i) \rangle \vert\notag\\
        =&\int d\Omega_{i_1}\int d\Omega_{i_2}\cdots\int d\Omega_{i_k}\notag\\
        &\quad\,\mu(\Omega_{i_1},\Omega_{i_2},\ldots,\Omega_{i_k})\prod_{i_s\in\mathcal{I}}\vert \langle \Delta_+A^{i_s}(\Omega_{i_s}) \rangle \vert\notag\\
        =&\prod_{i_s\in\mathcal{I}}\int d\Omega_{i_s}\mu_{i_s}(\Omega_{i_s})\vert \langle \Delta_+A^{i_s}(\Omega_{i_s}) \rangle \vert+\int d\Omega_{i_1}\cdots\int d\Omega_{i_k}\notag\\
        &\quad\left[\mu(\Omega_{i_1},\ldots,\Omega_{i_k})-\prod_{i_s\in\mathcal{I}}\mu_{i_s}(\Omega_{i_s})\right]\prod_{j_s\in\mathcal{I}}\vert \langle \Delta_+A^{j_s}(\Omega_{j_s}) \rangle \vert\notag\\
        \leq&\prod_{i_s\in\mathcal{I}}\int d\Omega_{i_s}\mu_{i_s}(\Omega_{i_s})\vert \langle \Delta_+A^{i_s}(\Omega_{i_s}) \rangle \vert+ \int d\Omega_{i_1}\cdots\int d\Omega_{i_k}\notag\\
        &\quad\,\,\vert\mu(\Omega_{i_1},\ldots,\Omega_{i_k})-\prod_{i_s\in\mathcal{I}}\mu_{i_s}(\Omega_{i_s})\vert\notag\\
        \leq&\prod_{i_s\in\mathcal{I}}\int d\Omega_{i_s}\mu_{i_s}(\Omega_{i_s})\vert \langle \Delta_+A^{i_s}(\Omega_{i_s}) \rangle \vert+\mathcal{M}_{n,k},
\end{align}
where the last line comes from taking the infimum.
Similarly, we have
\begin{equation}
    \vert J_{n,k}\vert\leq\prod_{i_s\in\mathcal{I}}\int d\Omega_{i_s}\mu_{i_s}(\Omega_{i_s})\vert \langle \Delta_-A^{i_s}(\Omega_{i_s}) \rangle \vert+\mathcal{M}_{n,k}.
\end{equation}
Using the Mahler inequality, we can conclude the proof:
    \begin{align}
        &{\left(\vert I_{n,k}\vert-\mathcal{M}_{n,k}\right)}^{\frac{1}{k}}+{\left(\vert J_{n,k}\vert-\mathcal{M}_{n,k}\right)}^{\frac{1}{k}}\notag\\
        \leq&\prod_{i_s\in\mathcal{I}}{\left[\int d\Omega_{i_s}\mu_{i_s}(\Omega_{i_s})(\vert\langle\Delta_+A^{i_s}(\Omega_{i_s})\rangle\vert + \vert\langle\Delta_-A^{i_s}(\Omega_{i_s})\rangle\vert) \right]}^{\frac{1}{k}}\notag\\
        \leq&1.
    \end{align}

\bibliographystyle{apsrev4-1}
\bibliography{BN_LSR}

@article{Braciard-2012a,
  title = {Bilocal versus nonbilocal correlations in entanglement-swapping experiments},
  author = {Branciard, Cyril and Rosset, Denis and Gisin, Nicolas and Pironio, Stefano},
  journal = {Phys. Rev. A},
  volume = {85},
  issue = {3},
  pages = {032119},
  numpages = {21},
  year = {2012},
  month = {Mar},
  publisher = {American Physical Society},
  doi = {10.1103/PhysRevA.85.032119},
  url = {https://link.aps.org/doi/10.1103/PhysRevA.85.032119}
}

@article{Peres-1999a,
  title = {All the Bell Inequalities},
  volume = {29},
  ISSN = {1572-9516},
  url = {http://dx.doi.org/10.1023/A:1018816310000},
  DOI = {10.1023/a:1018816310000},
  number = {4},
  journal = {Foundations of Physics},
  publisher = {Springer Science and Business Media LLC},
  author = {Peres,  Asher},
  year = {1999},
  month = Apr,
  pages = {589–614}
}

@article{deVicente-2017a,
  title = {Shared randomness and device-independent dimension witnessing},
  author = {de Vicente, Julio I.},
  journal = {Phys. Rev. A},
  volume = {95},
  issue = {1},
  pages = {012340},
  numpages = {11},
  year = {2017},
  month = {Jan},
  publisher = {American Physical Society},
  doi = {10.1103/PhysRevA.95.012340},
  url = {https://link.aps.org/doi/10.1103/PhysRevA.95.012340}
}

@article{Pal-2009a,
  title = {Concavity of the set of quantum probabilities for any given dimension},
  author = {P\'al, K\'aroly F. and V\'ertesi, Tam\'as},
  journal = {Phys. Rev. A},
  volume = {80},
  issue = {4},
  pages = {042114},
  numpages = {5},
  year = {2009},
  month = {Oct},
  publisher = {American Physical Society},
  doi = {10.1103/PhysRevA.80.042114},
  url = {https://link.aps.org/doi/10.1103/PhysRevA.80.042114}
}

@article{Sikora-2016a,
  title = {Minimum Dimension of a Hilbert Space Needed to Generate a Quantum Correlation},
  author = {Sikora, Jamie and Varvitsiotis, Antonios and Wei, Zhaohui},
  journal = {Phys. Rev. Lett.},
  volume = {117},
  issue = {6},
  pages = {060401},
  numpages = {5},
  year = {2016},
  month = {Aug},
  publisher = {American Physical Society},
  doi = {10.1103/PhysRevLett.117.060401},
  url = {https://link.aps.org/doi/10.1103/PhysRevLett.117.060401}
}

@article{Bell_1964_EPR,
  title = {On the Einstein Podolsky Rosen paradox},
  author = {Bell, J. S.},
  journal = {Physics Physique Fizika},
  volume = {1},
  issue = {3},
  pages = {195--200},
  numpages = {6},
  year = {1964},
  month = {Nov},
  publisher = {American Physical Society},
  doi = {10.1103/PhysicsPhysiqueFizika.1.195},
  url = {https://link.aps.org/doi/10.1103/PhysicsPhysiqueFizika.1.195}
}

@article{Braunstein_1988_info-theoretic-Bell-ineq,
  title = {Information-Theoretic Bell Inequalities},
  author = {Braunstein, Samuel L. and Caves, Carlton M.},
  journal = {Phys. Rev. Lett.},
  volume = {61},
  issue = {6},
  pages = {662--665},
  numpages = {0},
  year = {1988},
  month = {Aug},
  publisher = {American Physical Society},
  doi = {10.1103/PhysRevLett.61.662},
  url = {https://link.aps.org/doi/10.1103/PhysRevLett.61.662}
}

@article{Cerf_1997_entropic-Bell-ineq,
  title = {Entropic Bell inequalities},
  author = {Cerf, N. J. and Adami, C.},
  journal = {Phys. Rev. A},
  volume = {55},
  issue = {5},
  pages = {3371--3374},
  numpages = {0},
  year = {1997},
  month = {May},
  publisher = {American Physical Society},
  doi = {10.1103/PhysRevA.55.3371},
  url = {https://link.aps.org/doi/10.1103/PhysRevA.55.3371}
}

@article{Chaves-Fritz_2012_entropic,
  title = {Entropic approach to local realism and noncontextuality},
  author = {Chaves, Rafael and Fritz, Tobias},
  journal = {Phys. Rev. A},
  volume = {85},
  issue = {3},
  pages = {032113},
  numpages = {11},
  year = {2012},
  month = {Mar},
  publisher = {American Physical Society},
  doi = {10.1103/PhysRevA.85.032113},
  url = {https://link.aps.org/doi/10.1103/PhysRevA.85.032113}
}

@article{Chaves_2014_causal,
doi = {10.1088/1367-2630/16/4/043001},
url = {https://doi.org/10.1088/1367-2630/16/4/043001},
year = {2014},
month = {apr},
publisher = {IOP Publishing},
volume = {16},
number = {4},
pages = {043001},
author = {Chaves, Rafael and Luft, Lukas and Gross, David},
title = {Causal structures from entropic information: geometry and novel scenarios},
journal = {New Journal of Physics}
}

@article{Donohue_2015_idenfy-nonconvexity,
  title = {Identifying nonconvexity in the sets of limited-dimension quantum correlations},
  author = {Donohue, John Matthew and Wolfe, Elie},
  journal = {Phys. Rev. A},
  volume = {92},
  issue = {6},
  pages = {062120},
  numpages = {11},
  year = {2015},
  month = {Dec},
  publisher = {American Physical Society},
  doi = {10.1103/PhysRevA.92.062120},
  url = {https://link.aps.org/doi/10.1103/PhysRevA.92.062120}
}

@article{Fine_1982_HV,
  title = {Hidden Variables, Joint Probability, and the Bell Inequalities},
  author = {Fine, Arthur},
  journal = {Phys. Rev. Lett.},
  volume = {48},
  issue = {5},
  pages = {291--295},
  numpages = {0},
  year = {1982},
  month = {Feb},
  publisher = {American Physical Society},
  doi = {10.1103/PhysRevLett.48.291},
  url = {https://link.aps.org/doi/10.1103/PhysRevLett.48.291}
}

@article{Freedman_1972_Experiment-LHV,
  title = {Experimental Test of Local Hidden-Variable Theories},
  author = {Freedman, Stuart J. and Clauser, John F.},
  journal = {Phys. Rev. Lett.},
  volume = {28},
  issue = {14},
  pages = {938--941},
  numpages = {0},
  year = {1972},
  month = {Apr},
  publisher = {American Physical Society},
  doi = {10.1103/PhysRevLett.28.938},
  url = {https://link.aps.org/doi/10.1103/PhysRevLett.28.938}
}

@article{Fry_1976_Experimental-LHV,
  title = {Experimental Test of Local Hidden-Variable Theories},
  author = {Fry, Edward S. and Thompson, Randall C.},
  journal = {Phys. Rev. Lett.},
  volume = {37},
  issue = {8},
  pages = {465--468},
  numpages = {0},
  year = {1976},
  month = {Aug},
  publisher = {American Physical Society},
  doi = {10.1103/PhysRevLett.37.465},
  url = {https://link.aps.org/doi/10.1103/PhysRevLett.37.465}
}

@article{Aspect_1982_Experiment-EPR,
  title = {Experimental Realization of Einstein-Podolsky-Rosen-Bohm Gedankenexperiment: A New Violation of Bell's Inequalities},
  author = {Aspect, Alain and Grangier, Philippe and Roger, G\'erard},
  journal = {Phys. Rev. Lett.},
  volume = {49},
  issue = {2},
  pages = {91--94},
  numpages = {0},
  year = {1982},
  month = {Jul},
  publisher = {American Physical Society},
  doi = {10.1103/PhysRevLett.49.91},
  url = {https://link.aps.org/doi/10.1103/PhysRevLett.49.91}
}

@article{Weihs_1998_violations-of-Bell,
  title = {Violation of Bell's Inequality under Strict Einstein Locality Conditions},
  author = {Weihs, Gregor and Jennewein, Thomas and Simon, Christoph and Weinfurter, Harald and Zeilinger, Anton},
  journal = {Phys. Rev. Lett.},
  volume = {81},
  issue = {23},
  pages = {5039--5043},
  numpages = {0},
  year = {1998},
  month = {Dec},
  publisher = {American Physical Society},
  doi = {10.1103/PhysRevLett.81.5039},
  url = {https://link.aps.org/doi/10.1103/PhysRevLett.81.5039}
}

@article{Barrett_2005_Nonlocal-correlations,
  title = {Nonlocal correlations as an information-theoretic resource},
  author = {Barrett, Jonathan and Linden, Noah and Massar, Serge and Pironio, Stefano and Popescu, Sandu and Roberts, David},
  journal = {Phys. Rev. A},
  volume = {71},
  issue = {2},
  pages = {022101},
  numpages = {11},
  year = {2005},
  month = {Feb},
  publisher = {American Physical Society},
  doi = {10.1103/PhysRevA.71.022101},
  url = {https://link.aps.org/doi/10.1103/PhysRevA.71.022101}
}

@article{Pironio_2005_Lifting-Bell-ineq,
    author = {Pironio, Stefano},
    title = {Lifting Bell inequalities},
    journal = {Journal of Mathematical Physics},
    volume = {46},
    number = {6},
    pages = {062112},
    year = {2005},
    month = {06},
    issn = {0022-2488},
    doi = {10.1063/1.1928727},
    url = {https://doi.org/10.1063/1.1928727}
}

@article{Brunner_2014_Bell-nonlocality,
  title = {Bell nonlocality},
  author = {Brunner, Nicolas and Cavalcanti, Daniel and Pironio, Stefano and Scarani, Valerio and Wehner, Stephanie},
  journal = {Rev. Mod. Phys.},
  volume = {86},
  issue = {2},
  pages = {419--478},
  numpages = {60},
  year = {2014},
  month = {Apr},
  publisher = {American Physical Society},
  doi = {10.1103/RevModPhys.86.419},
  url = {https://link.aps.org/doi/10.1103/RevModPhys.86.419}
}

@misc{Dawei_2025_Quantum-Nonlocality-Latency,
      title={Quantum Nonlocality under Latency Constraints}, 
      author={Dawei Ding and Zhengfeng Ji and Pierre Pocreau and Mingze Xu and Xinyu Xu},
      year={2025},
      eprint={2510.26349},
      archivePrefix={arXiv},
      primaryClass={quant-ph},
      url={https://arxiv.org/abs/2510.26349}, 
}

@article{Acin_2007_DI,
  title = {Device-Independent Security of Quantum Cryptography against Collective Attacks},
  author = {Ac\'{\i}n, Antonio and Brunner, Nicolas and Gisin, Nicolas and Massar, Serge and Pironio, Stefano and Scarani, Valerio},
  journal = {Phys. Rev. Lett.},
  volume = {98},
  issue = {23},
  pages = {230501},
  numpages = {4},
  year = {2007},
  month = {Jun},
  publisher = {American Physical Society},
  doi = {10.1103/PhysRevLett.98.230501},
  url = {https://link.aps.org/doi/10.1103/PhysRevLett.98.230501}
}

@article{Pironio_2016_Focus-on-DI,
doi = {10.1088/1367-2630/18/10/100202},
url = {https://doi.org/10.1088/1367-2630/18/10/100202},
year = {2016},
month = {oct},
publisher = {IOP Publishing},
volume = {18},
number = {10},
pages = {100202},
author = {Pironio, S and Scarani, V and Vidick, T},
title = {Focus on device independent quantum information},
journal = {New Journal of Physics}
}

@article{Ekert_1991_QCrpto,
  title = {Quantum cryptography based on Bell's theorem},
  author = {Ekert, Artur K.},
  journal = {Phys. Rev. Lett.},
  volume = {67},
  issue = {6},
  pages = {661--663},
  numpages = {0},
  year = {1991},
  month = {Aug},
  publisher = {American Physical Society},
  doi = {10.1103/PhysRevLett.67.661},
  url = {https://link.aps.org/doi/10.1103/PhysRevLett.67.661}
}

@article{Acin_2006_From-Bell-to-QKD,
  title = {From Bell's Theorem to Secure Quantum Key Distribution},
  author = {Ac\'{\i}n, Antonio and Gisin, Nicolas and Masanes, Lluis},
  journal = {Phys. Rev. Lett.},
  volume = {97},
  issue = {12},
  pages = {120405},
  numpages = {4},
  year = {2006},
  month = {Sep},
  publisher = {American Physical Society},
  doi = {10.1103/PhysRevLett.97.120405},
  url = {https://link.aps.org/doi/10.1103/PhysRevLett.97.120405}
}

@article{Anders_2009_computational-power,
  title = {Computational Power of Correlations},
  author = {Anders, Janet and Browne, Dan E.},
  journal = {Phys. Rev. Lett.},
  volume = {102},
  issue = {5},
  pages = {050502},
  numpages = {4},
  year = {2009},
  month = {Feb},
  publisher = {American Physical Society},
  doi = {10.1103/PhysRevLett.102.050502},
  url = {https://link.aps.org/doi/10.1103/PhysRevLett.102.050502}
}

@article{Pozsgay_2017_cov-Bell-ineq,
  title = {Covariance Bell inequalities},
  author = {Pozsgay, Victor and Hirsch, Flavien and Branciard, Cyril and Brunner, Nicolas},
  journal = {Phys. Rev. A},
  volume = {96},
  issue = {6},
  pages = {062128},
  numpages = {13},
  year = {2017},
  month = {Dec},
  publisher = {American Physical Society},
  doi = {10.1103/PhysRevA.96.062128},
  url = {https://link.aps.org/doi/10.1103/PhysRevA.96.062128}
}

@article{Fritz_2012_beyond-Bell,
doi = {10.1088/1367-2630/14/10/103001},
url = {https://doi.org/10.1088/1367-2630/14/10/103001},
year = {2012},
month = {oct},
publisher = {IOP Publishing},
volume = {14},
number = {10},
pages = {103001},
author = {Fritz, Tobias},
title = {Beyond Bell's theorem: correlation scenarios},
journal = {New Journal of Physics}
}

@article{Elie_2019_inflation,
url = {https://doi.org/10.1515/jci-2017-0020},
title = {The Inflation Technique for Causal Inference with Latent Variables},
title = {},
author = {Elie Wolfe and Robert W. Spekkens and Tobias Fritz},
pages = {20170020},
volume = {7},
number = {2},
journal = {Journal of Causal Inference},
doi = {doi:10.1515/jci-2017-0020},
year = {2019},
lastchecked = {2026-01-20}
}

@article{Renou_2019_genuine-Q-triangle,
  title = {Genuine Quantum Nonlocality in the Triangle Network},
  author = {Renou, Marc-Olivier and B\"aumer, Elisa and Boreiri, Sadra and Brunner, Nicolas and Gisin, Nicolas and Beigi, Salman},
  journal = {Phys. Rev. Lett.},
  volume = {123},
  issue = {14},
  pages = {140401},
  numpages = {5},
  year = {2019},
  month = {Sep},
  publisher = {American Physical Society},
  doi = {10.1103/PhysRevLett.123.140401},
  url = {https://link.aps.org/doi/10.1103/PhysRevLett.123.140401}
}

@article{Renou_2022_Nonlocality-for-generic,
  title = {Nonlocality for Generic Networks},
  author = {Renou, Marc-Olivier and Beigi, Salman},
  journal = {Phys. Rev. Lett.},
  volume = {128},
  issue = {6},
  pages = {060401},
  numpages = {6},
  year = {2022},
  month = {Feb},
  publisher = {American Physical Society},
  doi = {10.1103/PhysRevLett.128.060401},
  url = {https://link.aps.org/doi/10.1103/PhysRevLett.128.060401}
}

@article{Pozas_2023_Proofs-network-Q-Nonlocality,
  title = {Proofs of Network Quantum Nonlocality in Continuous Families of Distributions},
  author = {Pozas-Kerstjens, Alejandro and Gisin, Nicolas and Renou, Marc-Olivier},
  journal = {Phys. Rev. Lett.},
  volume = {130},
  issue = {9},
  pages = {090201},
  numpages = {6},
  year = {2023},
  month = {Feb},
  publisher = {American Physical Society},
  doi = {10.1103/PhysRevLett.130.090201},
  url = {https://link.aps.org/doi/10.1103/PhysRevLett.130.090201}
}

@article{Luo_2018_computationally-efficient-Bell,
  title = {Computationally Efficient Nonlinear Bell Inequalities for Quantum Networks},
  author = {Luo, Ming-Xing},
  journal = {Phys. Rev. Lett.},
  volume = {120},
  issue = {14},
  pages = {140402},
  numpages = {6},
  year = {2018},
  month = {Apr},
  publisher = {American Physical Society},
  doi = {10.1103/PhysRevLett.120.140402},
  url = {https://link.aps.org/doi/10.1103/PhysRevLett.120.140402}
}

@article{Rosset_2016_nonlinear-Bell,
  title = {Nonlinear Bell Inequalities Tailored for Quantum Networks},
  author = {Rosset, Denis and Branciard, Cyril and Barnea, Tomer Jack and P\"utz, Gilles and Brunner, Nicolas and Gisin, Nicolas},
  journal = {Phys. Rev. Lett.},
  volume = {116},
  issue = {1},
  pages = {010403},
  numpages = {5},
  year = {2016},
  month = {Jan},
  publisher = {American Physical Society},
  doi = {10.1103/PhysRevLett.116.010403},
  url = {https://link.aps.org/doi/10.1103/PhysRevLett.116.010403}
}

@article{Tavakoli_2014_nonlocal-star,
  title = {Nonlocal correlations in the star-network configuration},
  author = {Tavakoli, Armin and Skrzypczyk, Paul and Cavalcanti, Daniel and Ac\'{\i}n, Antonio},
  journal = {Phys. Rev. A},
  volume = {90},
  issue = {6},
  pages = {062109},
  numpages = {12},
  year = {2014},
  month = {Dec},
  publisher = {American Physical Society},
  doi = {10.1103/PhysRevA.90.062109},
  url = {https://link.aps.org/doi/10.1103/PhysRevA.90.062109}
}

@article{Caratheodory_1907,
title = {Über den Variabilitätsbereich der Koeffizienten von Potenzreihen, die gegebene Werte nicht annehmen},
author = {Carathéodory, C},
journal = {Mathematische Annalen},
volume = {64},
issue = {1},
pages = {95},
year = {1907},
doi = {10.1007/BF01449883},
url = {https://doi.org/10.1007/BF01449883}
}

@ARTICLE{Fedotov_2003_Pinsker,
  author={Fedotov, A.A. and Harremoes, P. and Topsoe, F.},
  journal={IEEE Transactions on Information Theory}, 
  title={Refinements of Pinsker's inequality}, 
  year={2003},
  volume={49},
  number={6},
  pages={1491-1498},
  doi={10.1109/TIT.2003.811927}}

\end{document}